\documentclass[prb,twocolumn,preprintnumbers,amsmath,amssymb]{revtex4}% Physical Review B

\usepackage{graphicx}% Include figure files
\graphicspath{{fig/}}
\usepackage{dcolumn}% Align table columns on decimal point\varphi 
\usepackage{bm}% bold math
\usepackage{amsmath,amssymb}
\usepackage[T1]{fontenc}
\usepackage{textcomp}
\usepackage{color}
\begin{document}

\title{Asymmetric quantum shot noise in magnon transport
}

\author{Kouki Nakata,$^{1}$ Yuichi Ohnuma,$^{2}$ and Mamoru Matsuo$^{2}$}

\affiliation{$^1$Advanced Science Research Center, Japan Atomic Energy Agency, Tokai, Ibaraki 319-1195, Japan  \\
$^2$Kavli Institute for Theoretical Sciences, University of Chinese Academy of Sciences, Beijing, 100190, China
}

\date{\today}

%%%%%%%%%%%%%%%%%%%%%%%%%%%%%%%%%%%%%%%%%%%%%%%%%%%%%%%%%%%%%%%%%%%%%%%%%%%
%%%%%%%%%%%
\begin{abstract}
%%%%%%%%%%%
We study a frequency-dependent noise-to-current ratio for asymmetric, symmetric, and noncommutative current-noise in a ferromagnetic insulating junction, and extract quantum-mechanical properties of magnon transport at low temperatures. 
We demonstrate that the noncommutative noise, vanished in the dc-limit (i.e., a classical regime), increases monotonically as a function of frequency, 
and show that the noncommutative noise associated directly with quantum fluctuations of magnon currents breaks through the classical upper limit determined by the symmetric noise and realizes asymmetric quantum shot noise.
%%%%%%%%%%%%%%%%%%%%%%%%%%%%%%%%%%%%
Finally, we show that our theoretical predictions are within experimental reach with current device and measurement scheme
while exploiting low temperatures.
%%%%%%%%%%%%%%%%%%%%%%%%%%%%%%%%%%%%
Our work provides a platform toward experimental access to quantum fluctuations of magnon currents.
%%%%%%%%%
\end{abstract}
%%%%%%%%%
%%%%%%%%%%%%%%%%%%%%%%%%%%%%%%%%%%%%%%%%%%%%%%%%%%%%%%%%%%%%%%%%%%%%%%%%%%%%%%

\maketitle

%%%%%%%%%%%%%%%%%
%%%%%%%%%%%%%%%%%
\section{Introduction}
\label{sec:Intro}
%%%%%%%%%%%%%%%%%
%%%%%%%%%%%%%%%%%

\begin{figure}[h]
\begin{center}
\includegraphics[width=6cm,clip]{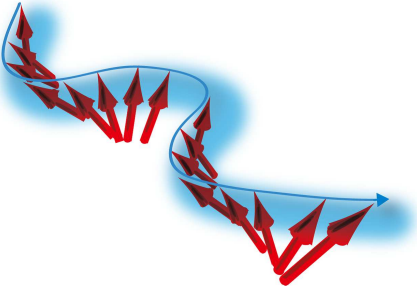}
\caption{
Schematic illustration of the spin-wave propagation with fluctuations.
The quantum version of spin-wave excitations, magnon, is the bosonic quasiparticle which carries magnetic momentum $\mu_{\rm{B}}$ in Bohr magneton units.
%the bosonic quasiparticle with a magnetic dipole moment $g \mu_{\rm{B}} {\mathbf{e}}_z$
%magnons which are chargeless bosonic quasiparticles with a magnetic dipole moment $g \mu_{\rm{B}} {\mathbf{e}}_z$  that serve as information carriers in Bohr magneton units $\mu_{\rm{B}}$.
%%%%%%%%%%%%%%%%%%%%%%%%%%%%%%%%%%%%%%%%%%%%%%
\label{fig:spinwave_arXiv}
}
\end{center}
\end{figure}

Spin-waves \cite{BlochMagnon} are the low-energy collective motion of localized spins and propagate via the exchange interaction in insulating magnets 
(Fig. \ref{fig:spinwave_arXiv}).
In the classical regime where spin operators can be identified with magnetization vectors being commutative,
the magnetization dynamics is described well by the Landau-Lifshitz-Gilbert (LLG) equation \cite{LLGtextbook,Gilbert}
that phenomenologically includes the damping term.
%%%%%%%%%%%%%%%%%%
Reflecting effects of thermal fluctuations as random fields, 
the stochastic LLG equation \cite{StochasticLLG} attempts to take into account spin fluctuations. 
%%%%%%%%%%%%%%%%%%%%%%%%%%%%%%%%%%%%%%%%

Thus the family of the LLG equation well describes essentially classical properties \cite{DLskyrmion} of the spin-wave propagation in the sense that the description holds only when 
the noncommutativity of spin operators ceases to work and spin operators are identified with the commutative magnetization vectors (e.g., macroscopic coherent spin precession \cite{demokritov,KKPD,ReviewMagnon}).
%%%%%%%%%%%%%%%%%%%%%%%%%%%%%%%%%%%%%%%%

%%%%%%%%%%%%%%%%%%%%%%%%
%%%%%%%%%%%%%%%%%%%%%%%%
\subsection{Theoretical background}
\label{subsec:Intro2}
%%%%%%%%%%%%%%%%%%%%%%%%
%%%%%%%%%%%%%%%%%%%%%%%%

A most promising strategy for theoretically exploring quantum-mechanical properties of the spin-wave propagation is to use the magnon-based description.
Via the Holstein-Primakoff expansion \cite{HP} respecting the noncommutativity of spin operators as the bosonic noncommutativity between creation and annihilation magnon operators, spin degrees of freedom are mapped into the magnon ones.

While in that sense magnons could be identified with the quantized version of spin-wave excitations, 
fully `quantum' transport phenomena associated directly with quantum fluctuations of `magnon currents' 
instead of `magnons' remain an open issue, to the best of our knowledge.
%%%%%%%%%%%%%%%%%%%%%%%%%%%%%%%%%%%%%%%%%%%%%%%%%%%%%%%%%%%%%%%%%
In this paper we provide a solution to this fundamental challenge in terms of magnonic current-noise, dubbed noncommutative noise for magnon currents 
(see Sec. \ref{subsec:Intro4} for details).
This is one of the purposes of this paper.

Note that the quantum-statistical properties of bosons and fermions are
fundamentally different, in particular in the low temperature regime where quantum effects dominate;
nonequilibrium noise plays a key role at such low temperatures in which thermal noise is suppressed,
and therefore 
it is natural to expect that shot noise properties vary from system to system, 
e.g., depending on whether bosons or fermions.
%%%%%%%%%%%%%%%%%%%%%%%%%%%%%%%%%%%%%%%%%%%%%%%%%%%%%%%%%%
While electron current-noise associated with quantum fluctuations has been well-understood \cite{NoiseRev}
for both theoretically \cite{Dahm,DLsuper-Poisson} and experimentally \cite{AsymmetricNoiseScience,AsymmetricNoisePRL},
it does not ensure that magnonic current-noise exhibits the same properties with electrons.
To clarify the properties of magnonic current-noise, in particular those of shot noise at low temperatures,
it requires careful and microscopic analysis taking into account the difference of quantum-statistical properties of bosons and fermions
(see Sec. \ref{subsec:ObservationAsymmetric} for details).

%%%%%%%%%%%%%%%%%%%%%%%%%%
%%%%%%%%%%%%%%%%%%%%%%%%%%
\subsection{Experimental background}
\label{subsec:Intro3}
%%%%%%%%%%%%%%%%%%%%%%%%%%
%%%%%%%%%%%%%%%%%%%%%%%%%%

The other purpose of this paper is as follows.
Recently, magnonic current-noise in an insulating ferromagnet has been measured as a function of frequency in Ref. [\onlinecite{magnonNoiseMeasurement}], see Fig. 3 (a) of Ref. [\onlinecite{magnonNoiseMeasurement}];
quite remarkably, magnonic current-noise is now a measurable physical quantity and 
the frequency-dependence is within experimental reach.
%%%%%%%%%%%%%%%%%%%%%%%%%%%%%%%%%%%%%%%%%%%%%%%%%%%%%%%%%%%%
However, while the frequency-dependence of electron current-noise has been well-understood \cite{NoiseRev}
for both theoretically \cite{Dahm,DLsuper-Poisson} and experimentally \cite{AsymmetricNoisePRL},
theoretical studies about the frequency-dependent magnonic current-noise have not yet enough \cite{magnonNoiseMeasurement,BalandinPrivate}.
Theoretical understanding of the frequency-dependence of current-noise carried by magnons, which are bosons instead of fermions, 
remains an open issue.

In this paper, 
we fill this gap by microscopically investigating frequency-dependent magnonic current-noise in an insulating magnet theoretically.
%%%%%%%%%%%%%%%%%%%%%%%%%%
%Within a junction model 
We thus theoretically provide some insights into experiment 
(see Sec. \ref{sec:TheoreticalInsight} for details).
This is the other purpose of this paper.
%%%%%%%%%%%%%%%%%%%%%%%%%%

%%%%%%%%%%%%%%%%%%%%%%%%%%%%%
%%%%%%%%%%%%%%%%%%%%%%%%%%%%%
%%%%%%%%%%%%%%%%%%%%%%%%%
%%%%%%%%%%%%%%%%%%%%%%%%%

We again remind the difference of the quantum-statistical properties of bosons and fermions,
in particular in the low temperature regime;
quantum effects dominate at low temperatures where
thermal noise is suppressed and nonequilibrium noise instead plays a significant role.
Therefore it is natural to expect that shot noise properties vary from system to system, 
e.g., depending on whether bosons or fermions,
and even whether it consists of spin or electric charge degrees of freedom;
%spin or electric charge;
%%%%%%%%%%%%%%%%%%%%%%%%%%%%%%%%%%%%%%%%%%%%%%%%%%%%%%%%%%
note that in quantum dots the dip near the zero-frequency regime due to the charging effects of the dot
has been predicted theoretically in Ref. [\onlinecite{DLsuper-Poisson}].
%%%%%%%%%%%%%%%%%%%%%%%%%%%%%%%%%%%%%%%%%%%%%%%%%%%%%%%%%%
While the frequency-dependence of electron current-noise thus has been well-understood \cite{NoiseRev}
for both theoretically and experimentally
(see, e.g., Refs. [\onlinecite{Dahm,DLsuper-Poisson}] for the theory and Ref. [\onlinecite{AsymmetricNoisePRL}] for the experiment),
it does not ensure that magnonic current-noise exhibits the same frequency-dependence with electrons.
%%%%%%%%%%%%%%%%%%%%%%%%%%%%%%%%%%%%%%%%%%%%%%%%%%%%%%%%%%
We need to microscopically analyze the frequency-dependence of magnonic current-noise
while taking into account the microscopic differences of systems, 
e.g., whether bosons or fermions, and whether it consists of spin or electric charge degrees of freedom, etc.
(see Sec. \ref{subsec:ObservationAsymmetric} for details).

%%%%%%%%%%%%%%%%%%%%%%%%%%%%%%%%
%%%%%%%%%%%%%%%%%%%%%%%%%%%%%%%%
\subsection{General properties of current-noise}
\label{subsec:Intro4}
%%%%%%%%%%%%%%%%%%%%%%%%%%%%%%%%
%%%%%%%%%%%%%%%%%%%%%%%%%%%%%%%%

%%%%%%%%%%%%%%%%%DLnoise2
%Explanation for each noise
%%%%%%%%%%%%%%%%%DLsuper-Poisson2%%%%%
Before going to the main subject, for readers' convenience 
it will be instructive to summarize general properties of current-noise.
%%%%%%%%%%%%%%%%%%%%%%%%%%%%%%
Via a (nonsymmetrized) current-current correlation function \cite{DLsuper-Poisson,AsymmetricNoiseScience,AsymmetricNoisePRL}, a current-noise  is defined by 
 $ {\cal{S}}^{\rm{asym}}(t, t')   \equiv   \langle {\hat{{\cal{I}}}}(t) {\hat{{\cal{I}}}}(t') \rangle    -    \langle {\hat{{\cal{I}}}}\rangle   \langle {\hat{{\cal{I}}}} \rangle$,
where $ \langle {\hat{{\cal{I}}}}\rangle  \equiv   {\cal{I}}  $ is the statistical average of a current operator $ {\hat{{\cal{I}}}} $.
Under the assumption 
$    \langle {\hat{{\cal{I}}}}(t) {\hat{{\cal{I}}}}(t') \rangle   \gg   \langle {\hat{{\cal{I}}}}\rangle   \langle {\hat{{\cal{I}}}} \rangle      $,
the asymmetric current-noise operator is introduced as $ {\hat{{\cal{S}}}}^{\rm{asym}}(t, t')   \equiv    {\hat{{\cal{I}}}}(t) {\hat{{\cal{I}}}}(t')     $, which can be rewritten by \footnote{
$ \{ {\hat{{\cal{I}}}}(t), {\hat{{\cal{I}}}}(t') \} \equiv  {\hat{{\cal{I}}}}(t) {\hat{{\cal{I}}}}(t') + {\hat{{\cal{I}}}}(t') {\hat{{\cal{I}}}}(t) $
while
$ [{\hat{{\cal{I}}}}(t), {\hat{{\cal{I}}}}(t')] \equiv   {\hat{{\cal{I}}}}(t) {\hat{{\cal{I}}}}(t') - {\hat{{\cal{I}}}}(t') {\hat{{\cal{I}}}}(t)  $.
}
\begin{eqnarray}
{\hat{{\cal{S}}}}^{\rm{asym}}(t, t') = \frac{ \{ {\hat{{\cal{I}}}}(t), {\hat{{\cal{I}}}}(t') \} }{2} +  \frac{  [{\hat{{\cal{I}}}}(t), {\hat{{\cal{I}}}}(t')]}{2}. 
\label{eqn:AsymNoise}                                       
\end{eqnarray}
This means that the asymmetric noise ${{\cal{S}}}^{\rm{asym}}(t, t')$ consists of two parts; 
the symmetric noise \cite{NoiseRev,DLnoise6,DLnoise5,DLnoise4,DLnoise3,DLsuper-Poisson2,KNmagnonNoiseJunction}
 ${{\cal{S}}}^{\rm{sym}}(t, t')$  and the noncommutative noise ${{\cal{S}}}^{\rm{nonc}}(t, t')$ whose operators are given by 
\begin{subequations}
\begin{eqnarray}
{\hat{{\cal{S}}}}^{\rm{sym}}(t, t') & \equiv  &   \frac{\{ {\hat{{\cal{I}}}}(t), {\hat{{\cal{I}}}}(t') \}}{2}
                                                                    =  {\hat{{\cal{S}}}}^{\rm{sym}}(t', t) ,    \label{eqn:SymNoise}    \\                                                                           
{\hat{{\cal{S}}}}^{\rm{nonc}}(t, t')  & \equiv  &   \frac{[{\hat{{\cal{I}}}}(t), {\hat{{\cal{I}}}}(t')]}{2}
                                                                      = -{\hat{{\cal{S}}}}^{\rm{nonc}}(t', t).       
\label{eqn:NoncNoise}                                                                          
\end{eqnarray}
\end{subequations}
For convenience, newly introducing a guiding operator for noise by ${\hat{{\cal{S}}}}(t, t') \equiv  {\hat{{\cal{S}}}}^{\rm{asym}}(t, t')/2 =  {\hat{{\cal{I}}}}(t) {\hat{{\cal{I}}}}(t')/2  $ and assuming the steady state in terms of time \cite{PeltierOhnuma} ${\hat{{\cal{S}}}}(t, t') = {\hat{{\cal{S}}}}(\delta {t}) $ with $\delta {t} \equiv  t-t'  $,
those noise operators are summarized in terms of $ {\hat{{\cal{S}}}}(\delta {t})   $;
\begin{subequations}
\begin{eqnarray}
{\hat{{\cal{S}}}}^{\rm{asym}}(\delta {t})   &=&     {\hat{{\cal{S}}}}^{\rm{sym}}(\delta {t}) +   {\hat{{\cal{S}}}}^{\rm{nonc}}(\delta {t})
                                                                             =    2 {\hat{{\cal{S}}}}(\delta {t}),   \label{eqn:Noise1}  \\
   {\hat{{\cal{S}}}}^{\rm{sym}}(\delta {t})   &=&   {\hat{{\cal{S}}}}^{\rm{sym}}(-\delta {t})
                                                                              = {\hat{{\cal{S}}}}(\delta {t}) +  {\hat{{\cal{S}}}}(- \delta {t}),   \label{eqn:Noise2}  \\
               {\hat{{\cal{S}}}}^{\rm{nonc}}(\delta {t})   &=&   -   {\hat{{\cal{S}}}}^{\rm{nonc}}(- \delta {t})
                                                                                           =  {\hat{{\cal{S}}}}(\delta {t}) -  {\hat{{\cal{S}}}}(- \delta {t}).       \label{eqn:Noise3}
\end{eqnarray}
\end{subequations}
Note that the asymmetric noise operator is not Hermitian
$ [{\hat{{\cal{S}}}}^{\rm{asym}}(\delta {t})]^{\dagger }/2 = [{\hat{{\cal{S}}}}(\delta {t})]^{\dagger } = {\hat{{\cal{S}}}}(- \delta {t}) \not =  {\hat{{\cal{S}}}}(\delta {t}) $,
while the noise spectrum 
being defined by $ {\cal{S}}(\Omega ) \equiv  \int d  (\delta {t}) {\rm{e}}^{i \Omega  \delta {t} }   {\cal{S}}(\delta {t})   $
takes a real value $  [{\cal{S}}(\Omega )]^{\ast } =   {\cal{S}}(\Omega ) $ and thereby
the asymmetric noise spectrum $ {\cal{S}}^{\rm{asym}}(\Omega ) =  2 {\cal{S}}(\Omega )  $ is observable.
%%%%%%%%%%%%%%%%%%%%%%%%%%%%%%%%%%%%%%%%%
In terms of ${\cal{S}}(\Omega )$, each noise spectrum is summarized as
\begin{subequations}
\begin{eqnarray}
{\cal{S}}^{\rm{asym}}(\Omega )   &=& {\cal{S}}^{\rm{sym}}(\Omega )+ {\cal{S}}^{\rm{nonc}}(\Omega )   =  2 {\cal{S}}(\Omega ), 
            \label{eqn:Noise1b}  \\
   {\cal{S}}^{\rm{sym}}(\Omega )   &=&   {\cal{S}}^{\rm{sym}}(-\Omega )= {\cal{S}}(\Omega ) +  {\cal{S}}(-\Omega ), 
                \label{eqn:Noise2b}  \\
               {\cal{S}}^{\rm{nonc}}(\Omega )   &=&     -    {\cal{S}}^{\rm{nonc}}(-\Omega ) =   {\cal{S}}(\Omega ) -   {\cal{S}}(-\Omega ).  
                     \label{eqn:Noise3b}
\end{eqnarray}
\end{subequations}
%%%%%%%%%%%%%%%%%%%%%%%%%%%%%%%%%%%%%%%%%

In the classical regime where the noncommutativity of spin operators ceases to work and spin operators are identified with 
commutative magnetization vectors,
the noncommutative noise vanishes  ${{\cal{S}}}^{\rm{nonc}}(\delta  t)=0$.
%%%%%%%%%%%%%%%%%%%%%%%%%%%%%%%%%%%%%%%%%%%%%
Therefore making use of the noncommutative noise spectrum $ {\cal{S}}^{\rm{nonc}}(\Omega )$ and in particular focusing on the contribution to the asymmetric noise spectrum $  {\cal{S}}^{\rm{asym}}(\Omega )   = {\cal{S}}^{\rm{sym}}(\Omega )+ {\cal{S}}^{\rm{nonc}}(\Omega )   $,
we explore the quantum-mechanical properties of magnon transport associated with the noncommutativity of the current operator.
%%%%%%%%%%%%%%%
Note that as seen in the Heisenberg's uncertainty relation \cite{HeisenbergUncertain,HeisenbergUncertain2,HeisenbergUncertain3}, the noncommutativity of operators lies at the heart of quantum mechanics.
Thereby in terms of the noncommutative noise spectrum associated directly with quantum fluctuations of magnon currents, 
we find a fully `quantum' transport phenomenon of magnons. 
%which cannot be obtained by the LLG equation-based description.
%%%%%%%%%%%%%%%%%%%%%%%%%%%%%%%%%%%%%%%%%%%%%

%Significance of \Omega%%%%%%%%%%%%%%%%%%%%%%%%%%%%%%%%%%%%%%%%%%%%
We remark that since $ {\cal{S}}^{\rm{sym}}(-\Omega ) = {\cal{S}}^{\rm{sym}}(\Omega )  $ while 
${\cal{S}}^{\rm{nonc}}(-\Omega ) = - {\cal{S}}^{\rm{nonc}}(\Omega )   $,
the asymmetry of the asymmetric noise spectrum ${\cal{S}}^{\rm{asym}}(- \Omega )  \not=  {\cal{S}}^{\rm{asym}}(\Omega ) $ as a function of $\Omega $ arises from the noncommutative noise spectrum ${\cal{S}}^{\rm{nonc}}(\Omega )$.
%%%%%%%%%%%%%%%%%%%%%%%%%%%%%%%%%%%%%%%%%%%%%
Moreover, in the dc-limit $ \Omega =0$ the noncommutative noise spectrum vanishes 
\begin{eqnarray}
         {\cal{S}}^{\rm{nonc}}(\Omega =0) = 0         \label{eqn:noncNoise0}                
\end{eqnarray}
and the asymmetric one becomes identical to the symmetric one as
\begin{eqnarray}
  {\cal{S}}^{\rm{asym}}(\Omega =0) = {\cal{S}}^{\rm{sym}}(\Omega =0).           \label{eqn:Noise0}                           
\end{eqnarray}
%%%%%%%%%%%%%%%%%%%%%%%%%%%%%%%
These indicate that the frequency $\Omega $ is a good parameter to characterize quantum-mechanical features of the noise spectrum.

Those are the general properties of current-noise.

%%%%%%%%%%%%%%%%%%%%%%%%%%%%%%%%%%%%
%%%%%%%%%%%%%%%%%%%%%%%%%%%%%%%%%%%%
This paper is organized as follows.
%%%%%%%%%%%%%%%%%%%%%%%%%%%%%%%%%%%%
%%%%%%%%%%%%%%%%%%%%%%%%%%%%%%%%%%%%
In Sec. \ref{sec:system} we introduce the model system for magnon transport with fluctuations across the junction interface of ferromagnetic insulators.
In Sec. \ref{sec:3} we describe asymmetric, symmetric, and noncommutative noise for magnonic spin and heat currents,
and derive the frequency-dependent noise spectrum. 
In Sec. \ref{sec:4} we determine the frequency-dependent noise-to-current ratio and demonstrate that the noncommutative noise associated directly with quantum fluctuations of magnon currents breaks through the classical limit and realizes asymmetric quantum shot noise.
%%%%%%%%%%%%%%%%%%%%%%%%%%%%%%%%%%%%
In Sec. \ref{sec:TheoreticalInsight} we provide some insights into experiment.
%%%%%%%%%%%%%%%%%%%%%%%%%%%%%%%%%%%%
Finally, we give some conclusions in Sec. \ref{sec:conclusion},
and remark open issues in Sec. \ref{sec:discussion}.

%%%%%%%%%%%%%%%%%
%%%%%%%%%%%%%%%%%
\section{System}
\label{sec:system}
%%%%%%%%%%%%%%%%%
%%%%%%%%%%%%%%%%%

As a platform to extract quantum-mechanical properties of magnon transport associated with the noncommutativity of the current operator, 
we consider the simplest model, a three-dimensional ferromagnetic insulating junction \cite{KNmagnonNoiseJunction} 
formed by two ferromagnetic insulators (FIs) aligned along the $x$-direction,
and focus on spin and heat currents carried by magnons across the junction interface embedded in the $yz$-plane \cite{magnonWF}.
%%%%%%%%%%%%%%%%%%%%%%%%%%%%%%%%%%%%%KNmagnonNoiseJunction
There exists in general a finite overlap of the wave functions between the spins ${\bf S}_{\Gamma_{\rm{R}}}$ and ${\bf S}_{\Gamma_{\rm{L}}}$ of length $S \gg  1$ located at the boundary of the right and left FI, respectively.
This results in a finite exchange interaction $ J_{\rm{ex}} > 0$ between the two FIs.
%%%%%%%%%%%%%%%%%%%%%%%%%%%%%%%%%%%%%%%%%%%%%%%
%As a minimal model (i.e., setup) to extract quantum-mechanical properties of magnon transport associated with the noncommutativity of the current operator,
We assume that the exchange interaction $ J_{\rm{ex}} $  is weak \footnote{
We expect that such a weak exchange interaction between the boundary spins may be experimentally realized by placing an ultrathin film of a nonmagnetic material between the two FIs since the exchange interaction arises essentially from a finite overlap of the wave functions between the boundary spins.
}
compared with the one between the nearest neighboring spins in each FI and may be described by the Hamiltonian
${\cal{H}}_{\rm{ex}}  = -J_{\rm{ex}} \sum_{\langle \Gamma_{\rm{L}} \Gamma_{\rm{R}} \rangle} {\bf S}_{\Gamma_{\rm{L}}} \cdot {\bf S}_{\Gamma_{\rm{R}}}$.
%%%%%%%%%%%%%%%%%%%%%%%%%%%%%%%
The spins in each three-dimensional FI are described by the Heisenberg spin Hamiltonian on cubic lattice with the Zeeman term
 $g \mu _{\rm{B}}B_{\rm{L(R)}}$ determined by the applied magnetic field $ B_{\rm{L(R)}} \equiv B $ to the left (right) FI.
 Including the $g$-factor, the left FI is identical to the right one except the temperature, $ T_{\rm{L}} \equiv  T  $ and $ T_{\rm{R}} \equiv  T + \Delta T  $, respectively.
%%%%%%%%%%%%%%%%%%%%%%%%%%%%

Applying strong magnetic field enough, we assume magnetic order along the magnetic field, which defines the $z$-direction.
%%%%%%%%%%%%%%%%%%%%%%%%%%%%
Via the Holstein-Primakoff expansion \cite{HP} to leading order, spin degrees of freedom are mapped into the magnon ones
and the Hamiltonian reduces to
$ {\cal{H}}_{\rm{ex}} =  - J_{{\rm{ex}}} S \sum_{{\mathbf{k}}_{\perp }}  \sum_{k_x, k_x^{\prime}}
a_{{\rm{L}},  {\mathbf{k}}}  a^{\dagger }_{{\rm{R}}, {\mathbf{k}}^{\prime}} + {\rm{H. c.}}   $,
where the bosonic operator $a_{\rm{R/L}}^{\dagger}$ ($a_{\rm{R/L}}$) creates (annihilates) the magnon at the boundary of the right/left FI that possesses the momentum ${\mathbf{k}}^{\prime}=(k_x^{\prime}, k_y, k_z)$ and ${\mathbf{k}}=(k_x, k_y, k_z)$, respectively, with ${\mathbf{k}}_{\perp }=(0, k_y, k_z)  $.
%%%%%%%%%%%%%%%%%%%%%%%%%%%%%%%

%%%%%%%%%%%%%%%%%%%%%%%%%%%%%%%%%
%%%%%%%%%%%%%%%%%%%%%%%%%%%%%%%%%
\section{Noise spectrum}
\label{sec:3}
%%%%%%%%%%%%%%%%%%%%%%%%%%%%%%%%%
%%%%%%%%%%%%%%%%%%%%%%%%%%%%%%%%%

The tunneling Hamiltonian $ {\cal{H}}_{\rm{ex}}$ produces the time-evolution of magnon operators for each FI, and generates magnonic spin and heat currents across the junction interface.
Using the Heisenberg equation of motion in terms of the left FI, 
we define the magnonic spin and heat current operators \cite{PeltierOhnuma},
$ \hat{{\cal{I}}}_{\rm{m}} (t)$ and  $ \hat{{\cal{I}}}_{Q} (t)$, respectively, flowing across the junction interface from the right FI to the left one
(see Ref. [\onlinecite{KNmagnonNoiseJunction}] for details);
%%%%%%%%%%%%%%%%%%%%%%%%%%%%%%%%%%%%%%%%%%%%%%%%
$ {\hat{{\cal{I}}}}_{\rm{m}}  (t) 
\equiv  g \mu _{\rm{B}} \partial _{t} N_{\rm{L}}(t) 
= - i  g \mu _{\rm{B}} (J_{\rm{ex}}S/\hbar ) \sum_{{\mathbf{k}}, k_x^{\prime}}
a_{{\rm{L}},{\mathbf{k}}}(t)  a^{\dagger }_{{\rm{R}},{\mathbf{k}}^{\prime}}(t) + {\rm{H. c.}}$,
%%%%%%%%%%%%%%%%%%%%%%%%%%%%%%%%
$ {\hat{{\cal{I}}}}_Q  (t) 
\equiv \partial _{t} {\cal{H}}_{\rm{L}}
= J_{\rm{ex}}S  \sum_{{\mathbf{k}}, k_x^{\prime}}
 [\partial _{t}^{\rm{L}} a_{{\rm{L}},{\mathbf{k}}}(t)]  a^{\dagger }_{{\rm{R}},{\mathbf{k}}^{\prime}}(t) + {\rm{H. c.}}$,
 where $   N_{\rm{L}}(t) \equiv   \sum_{\mathbf{q}} a_{{\rm{L}},{\mathbf{q}}}^{\dagger }(t) a_{{\rm{L}},{\mathbf{q}}}(t)$
is the magnon number operator of the left FI described by the Hamiltonian $ {\cal{H}}_{\rm{L}}  $
and the time-derivative $ \partial _{t}^{\rm{L}}  $ which works on the magnon operators solely for the left FI.
%%%%%%%%%%%%%%%%%%%%%%%%%%%%%%%%
%%%%%%%%%%%%%%%%%%%%%%%%%%%%%%%%

A straightforward perturbative calculation up to ${\cal{O}} (J_{\rm{ex}}^2)$ based on the Schwinger-Keldysh formalism \cite{Schwinger,Schwinger2,Keldysh,haug,tatara,kita} gives the statistical average of those magnonic currents $ \langle  {\hat{{\cal{I}}}}_{{\rm{m}}(Q)} \rangle  \equiv  {\cal{I}}_{{\rm{m}}(Q)} $ by 
(see Ref. [\onlinecite{KNmagnonNoiseJunction}] for details)
%%%%%%%%%%%%%%%%%%%%%%%%%%%%%%%%%%%%%%%
$ {\cal{I}}_{\rm{m}} =  [(J_{\rm{ex}}S)^2/2 \pi]   \sum_{{\mathbf{k}}, k_x^{\prime}}  \int  d\omega  g \mu _{\rm{B}}
    (G_{{\rm{L}}, {\mathbf{k}}, \omega }^{<}   G_{{\rm{R}}, {\mathbf{k}}^{\prime}, \omega  }^{>}  
   - G_{{\rm{L}}, {\mathbf{k}}, \omega }^{>}   G_{{\rm{R}}, {\mathbf{k}}^{\prime}, \omega   }^{<})  = {\cal{O}} (J_{\rm{ex}}^2)  $,
%%%%%%%%%%%%%%%%%%%%%%%%%%%%%%%%%%%%%%%
$ {\cal{I}}_{Q} =  [(J_{\rm{ex}}S)^2/2 \pi]   \sum_{{\mathbf{k}}, k_x^{\prime}}  \int  d\omega   \hbar \omega  
   (G_{{\rm{L}}, {\mathbf{k}}, \omega }^{<}   G_{{\rm{R}}, {\mathbf{k}}^{\prime}, \omega  }^{>}  
  - G_{{\rm{L}}, {\mathbf{k}}, \omega }^{>}   G_{{\rm{R}}, {\mathbf{k}}^{\prime}, \omega   }^{<}) = {\cal{O}} (J_{\rm{ex}}^2)  $,
%%%%%%%%%%%%%%%%%%%%%%%%%%%%%%%%%%%%%%%
where $   G_{{\rm{L}}, {\mathbf{k}}, \omega }^{>(<)}  $ and $ G_{{\rm{R}}, {\mathbf{k}}^{\prime}, \omega }^{>(<)}  $
are the bosonic greater (lesser) Green's functions of magnons for the left and right FIs, respectively.
%%%%%%%%%%%%%%%%%%%%%%%%%%%%%%%
Since  
\begin{eqnarray}
 \langle {\hat{{\cal{I}}}}_{{\rm{m}}(Q)}\rangle   \langle {\hat{{\cal{I}}}}_{{\rm{m}}(Q)} \rangle = {\cal{O}} (J_{\rm{ex}}^4) 
\end{eqnarray}
while 
$  \langle {\hat{{\cal{I}}}}_{{\rm{m}}(Q)}(t) {\hat{{\cal{I}}}}_{{\rm{m}}(Q)}(t') \rangle  = {\cal{O}} (J_{\rm{ex}}^2)   $ as seen below [Eqs. (\ref{eqn:SmOmega}) and (\ref{eqn:SQOmega})], 
the asymmetric spin (heat) current-noise of magnons in the junction becomes 
\begin{subequations}
\begin{eqnarray}
  {\cal{S}}^{\rm{asym}}_{{\rm{m}}(Q)}(t, t')  & \equiv  &  \langle {\hat{{\cal{I}}}}_{{\rm{m}}(Q)}(t) {\hat{{\cal{I}}}}_{{\rm{m}}(Q)}(t') \rangle    
-    \langle {\hat{{\cal{I}}}}_{{\rm{m}}(Q)}\rangle   \langle {\hat{{\cal{I}}}}_{{\rm{m}}(Q)} \rangle         \    \    \    \    \       \\
&=& \langle  {\hat{{\cal{I}}}}_{{\rm{m}}(Q)}(t) {\hat{{\cal{I}}}}_{{\rm{m}}(Q)}(t')  \rangle +  {\cal{O}} (J_{\rm{ex}}^4). 
\end{eqnarray}
\end{subequations}
%$  {\cal{S}}^{\rm{asym}}_{{\rm{m}}(Q)}(t, t')   = \langle  {\hat{{\cal{I}}}}_{{\rm{m}}(Q)}(t) {\hat{{\cal{I}}}}_{{\rm{m}}(Q)}(t')  \rangle +  {\cal{O}} (J_{\rm{ex}}^4)  $
%%%%%%%%%%%%%%%%%%%%%%%%%%%%%%%%%
Thereby the operator is given by
\begin{eqnarray}
{\hat{{\cal{S}}}}^{\rm{asym}}_{{\rm{m}}(Q)}(t, t')   =    {\hat{{\cal{I}}}}_{{\rm{m}}(Q)}(t) {\hat{{\cal{I}}}}_{{\rm{m}}(Q)}(t')
\label{eqn:asymNoiseSpinHeat}  
\end{eqnarray}
in the weak exchange-coupling regime.
%%%%%%%%%%%%%%%%%%%%%%%%%%%%%%%%%

For convenience, as a guiding operator for noise we introduce newly a spin (heat) current-noise operator by
$  {\hat{{\cal{S}}}}_{{\rm{m}}(Q)}(t, t')   \equiv   {\hat{{\cal{S}}}}^{\rm{asym}}_{{\rm{m}}(Q)}(t, t') /2  =    {\hat{{\cal{I}}}}_{{\rm{m}}(Q)}(t) {\hat{{\cal{I}}}}_{{\rm{m}}(Q)}(t')/2  $
and assume the steady state in terms of time \cite{PeltierOhnuma} ${\hat{{\cal{S}}}}_{{\rm{m}}(Q)}(t, t') = {\hat{{\cal{S}}}}_{{\rm{m}}(Q)}(\delta {t}) $.
%%%%%%%%%%%%%%%%%%%%%%%%%%%%%%%%%%%%%%%%%%%%%%%%%%%%%
Using $ {\hat{{\cal{S}}}}_{{\rm{m}}(Q)}(\delta {t})   $, the asymmetric noise operator is characterized by 
  ${\hat{{\cal{S}}}}^{\rm{asym}}_{{\rm{m}}(Q)}(\delta {t})   =  2 {\hat{{\cal{S}}}}_{{\rm{m}}(Q)}(\delta {t})
           =    {\hat{{\cal{S}}}}^{\rm{sym}}_{{\rm{m}}(Q)}(\delta {t}) +   {\hat{{\cal{S}}}}^{\rm{nonc}}_{{\rm{m}}(Q)}(\delta {t})$,
where the symmetric and noncommutative spin (heat) current-operators, 
$  {\hat{{\cal{S}}}}^{\rm{sym}}_{{\rm{m}}(Q)}(t, t')  $ and $  {\hat{{\cal{S}}}}^{\rm{nonc}}_{{\rm{m}}(Q)}(t, t')   $, respectively, are defined by
\begin{subequations}
\begin{eqnarray}
 {\hat{{\cal{S}}}}^{\rm{sym}}_{{\rm{m}}(Q)}(t, t')  & \equiv &   {\{ {\hat{{\cal{I}}}}_{{\rm{m}}(Q)}(t), {\hat{{\cal{I}}}}_{{\rm{m}}(Q)}(t') \}}/{2}, 
 \label{eqn:symNoiseSpinHeat}      \\     
{\hat{{\cal{S}}}}^{\rm{nonc}}_{{\rm{m}}(Q)}(t, t')   & \equiv &     {[{\hat{{\cal{I}}}}_{{\rm{m}}(Q)}(t), {\hat{{\cal{I}}}}_{{\rm{m}}(Q)}(t')]}/{2}, 
\label{eqn:noncNoiseSpinHeat} 
\end{eqnarray}
\end{subequations}
and those satisfy
%%%%%%%%%%%%%%%%%%%%%%%%%%%%%%%%%%
 ${\hat{{\cal{S}}}}^{\rm{sym}}_{{\rm{m}}(Q)}(\delta {t})   =   {\hat{{\cal{S}}}}_{{\rm{m}}(Q)}(\delta {t}) +  {\hat{{\cal{S}}}}_{{\rm{m}}(Q)}(- \delta {t})$ 
and
%%%%%%%%%%%%%%%%%%%%%%%%%%%%%%%%%%%%%%%%%%%%
$ {\hat{{\cal{S}}}}^{\rm{nonc}}_{{\rm{m}}(Q)}(\delta {t})   =        {\hat{{\cal{S}}}}_{{\rm{m}}(Q)}(\delta {t}) -  {\hat{{\cal{S}}}}_{{\rm{m}}(Q)}(- \delta {t})$
in the steady state.
%%%%%%%%%%%%%%%%%%%%%%%%%%%%%%%%%%%%%%%%%%%%%%%%%
%%%%%%%%%%%%%%%%%%%%%%%%%%%%%%%%%%%%%%%%%%%%%%%%%
In terms of the noise spectrum being defined by
\begin{eqnarray}
 {\cal{S}}_{{\rm{m}}(Q)}(\Omega ) \equiv  \int d  (\delta {t}) {\rm{e}}^{i \Omega  \delta {t} }   {\cal{S}}_{{\rm{m}}(Q)}(\delta {t}),
 \label{eqn:DefNoiseSpectrumKKK} 
\end{eqnarray}
each noise spectrum is summarized as
${\cal{S}}^{\rm{asym}}_{{\rm{m}}(Q)}(\Omega )   =2 {\cal{S}}_{{\rm{m}}(Q)}(\Omega )
  =  {\cal{S}}^{\rm{sym}}_{{\rm{m}}(Q)}(\Omega )+ {\cal{S}}^{\rm{nonc}}_{{\rm{m}}(Q)}(\Omega )$,   
%%%%%%%%%%%%%%%%%%%%%%%%%%%%%%%%%%%%%%%%%%%%%%%%%%%%%%%%%%%%%%
 $  {\cal{S}}^{\rm{sym}}_{{\rm{m}}(Q)}(\Omega )   =   {\cal{S}}_{{\rm{m}}(Q)}(\Omega ) +  {\cal{S}}_{{\rm{m}}(Q)}(-\Omega )$,  
%%%%%%%%%%%%%%%%%%%%%%%%%%%%%%%%%%%%%%%%%%%%%%%%%%%%%%%%%%%%%%
  ${\cal{S}}^{\rm{nonc}}_{{\rm{m}}(Q)}(\Omega )   =        {\cal{S}}_{{\rm{m}}(Q)}(\Omega ) -   {\cal{S}}_{{\rm{m}}(Q)}(-\Omega )$.
%%%%%%%%%%%%%%%%%%%%%%%%%%%%%%%%%%%%%%%%%%
%%%%%%%%%%%%%%%%%%%%%%%%%%%%%%%%%%%%%%%%%%

The asymmetric, symmetric, and noncommutative current-noise spectrum, 
$  {\cal{S}}^{\rm{asym}}_{{\rm{m}}(Q)}(\Omega )$,
$  {\cal{S}}^{\rm{sym}}_{{\rm{m}}(Q)}(\Omega )  $,
and $ {\cal{S}}^{\rm{nonc}}_{{\rm{m}}(Q)}(\Omega ) $, respectively, 
consists of $ {\cal{S}}_{{\rm{m}}(Q)}(\Omega )  $.
A straightforward perturbative calculation up to ${\cal{O}} (J_{\rm{ex}}^2)$ based on the Schwinger-Keldysh formalism \cite{Schwinger,Schwinger2,Keldysh,haug,tatara,kita} provides 
(see Ref. [\onlinecite{KNmagnonNoiseJunction}] for details)
\begin{subequations}
\begin{eqnarray}
%%%%%%%%%%%%%%%%%%%%%%%%%%%%%%%%%%
{\cal{S}}_{\rm{m}}(\Omega ) &=&  -  \frac{(J_{\rm{ex}}S)^2}{4 \pi}   \sum_{{\mathbf{k}}, k_x^{\prime}}  \int  d\omega  (g \mu _{\rm{B}})^2   \nonumber   \\
& \times &  (G_{{\rm{L}}, {\mathbf{k}}, \omega }^{>}   G_{{\rm{R}}, {\mathbf{k}}^{\prime}, \omega - \Omega }^{<}  
 + G_{{\rm{L}}, {\mathbf{k}}, \omega }^{<}   G_{{\rm{R}}, {\mathbf{k}}^{\prime}, \omega + \Omega  }^{>}),
\    \   \     \       \label{eqn:SmOmega}  \\
%%%%%%%%%%%%%%%%%%%%%%%%%%%%%%%%%%
%%%%%%%%%%%%%%%%%%%%%%%%%%%%%%%%%%
{\cal{S}}_{Q}(\Omega ) &=&  -  \frac{(J_{\rm{ex}}S)^2}{4 \pi}   \sum_{{\mathbf{k}}, k_x^{\prime}}  \int  d\omega  (\hbar \omega )^2   \nonumber   \\
& \times &  (G_{{\rm{L}}, {\mathbf{k}}, \omega }^{>}   G_{{\rm{R}}, {\mathbf{k}}^{\prime}, \omega - \Omega }^{<}  
 + G_{{\rm{L}}, {\mathbf{k}}, \omega }^{<}   G_{{\rm{R}}, {\mathbf{k}}^{\prime}, \omega + \Omega  }^{>}).
\   \    \     \    \label{eqn:SQOmega}
%%%%%%%%%%%%%%%%%%%%%%%%%%%%%%%%%%
\end{eqnarray}
\end{subequations}
%%%%%%%%%%%%%%%%%%%%%%%%%
The bosonic greater and lesser Green's functions are rewritten in terms of the bosonic retarded Green's function \cite{AGD}
$G^{\rm{r}}_{{\mathbf{k}}, \omega }= (\hbar \omega  -  \omega_{{\mathbf{k}}} +i \eta \hbar  \omega  )^{-1} $ 
by 
$ G^{>}_{{\mathbf{k}}, \omega } = 2 i [1+n(\omega )] {\rm{Im}}  G^{\rm{r}}_{{\mathbf{k}}, \omega }$ and 
$ G^{<}_{{\mathbf{k}}, \omega } = 2 i n(\omega ) {\rm{Im}}  G^{\rm{r}}_{{\mathbf{k}}, \omega }$, respectively,
where the Bose-distribution function $ n(\omega )=({\rm{e}}^{\beta \hbar \omega }-1)^{-1}  $, $\beta  = 1/k_{\rm{B}}T  $,
the energy dispersion of magnons $  \omega_{{\mathbf{k}}}$, and $\eta $ ($  \eta  \geq  0 $) is a constant that characterizes magnon lifetime.
%%%%%%%%%%%%%%$\eta $ is a constant that characterizes the magnon lifetime $\tau$ by $ 1/2 \tau  =  \eta \omega  $.
Since throughout this paper we focus on sufficiently low temperatures (i.e., magnonic shot noise regime) in which effects of magnon-magnon and magnon-phonon interactions become negligibly small \cite{Tmagnonphonon,adachiphonon,magnonWF}, we assume that the magnon lifetime is temperature-independent; nonmagnetic impurity scatterings account for the temperature-independent lifetime and $\eta $.
%%%%%%%%%%%%%%%%%%%%
%%%%%%%%%%%%%%%%%%%%
Thus phenomenologically taking into account the effects of nonmagnetic impurity scatterings (i.e., disorder effects), we obtain the frequency-dependent noise spectrum
\begin{subequations}
\begin{eqnarray}
%%%%%%%%%%%%%%%%%%%%%%%%%%%%%%%%%%
{\cal{S}}_{\rm{m}}(\Omega ) &=&   \frac{(J_{\rm{ex}}S)^2}{\pi}   \sum_{{\mathbf{k}}, k_x^{\prime}}  \int d \omega   (g \mu _{\rm{B}})^2  
                                                                                                                                                                                                \label{eqn:Sm2Omega}      \\
&\times & {\Big{\{ }}{\rm{Im}} G_{{\rm{L}}, {\mathbf{k}}, \omega }^{\rm{r}}   {\rm{Im}} G_{{\rm{R}}, {\mathbf{k}}^{\prime}, \omega -\Omega  }^{\rm{r}} 
   [1+n_{\rm{L}}(\omega )] n_{\rm{R}}(\omega -\Omega ) \nonumber  \\
&+& {\rm{Im}} G_{{\rm{L}}, {\mathbf{k}}, \omega }^{\rm{r}}   {\rm{Im}} G_{{\rm{R}}, {\mathbf{k}}^{\prime}, \omega +\Omega  }^{\rm{r}} 
     n_{\rm{L}}(\omega ) [1+n_{\rm{R}}(\omega +\Omega )] {\Big{\} }},  \nonumber \\
%%%%%%%%%%%%%%%%%%%%%%%%%%%%%%%%%%
%%%%%%%%%%%%%%%%%%%%%%%%%%%%%%%%%%
{\cal{S}}_{Q}(\Omega ) &=&   \frac{(J_{\rm{ex}}S)^2}{\pi}   \sum_{{\mathbf{k}}, k_x^{\prime}}  \int d \omega   (\hbar \omega  )^2 
                                                                                                                                                                \label{eqn:SQ2Omega}  \\
&\times & {\Big{\{ }}{\rm{Im}} G_{{\rm{L}}, {\mathbf{k}}, \omega }^{\rm{r}}   {\rm{Im}} G_{{\rm{R}}, {\mathbf{k}}^{\prime}, \omega -\Omega  }^{\rm{r}} 
   [1+n_{\rm{L}}(\omega )] n_{\rm{R}}(\omega -\Omega ) \nonumber  \\
&+& {\rm{Im}} G_{{\rm{L}}, {\mathbf{k}}, \omega }^{\rm{r}}   {\rm{Im}} G_{{\rm{R}}, {\mathbf{k}}^{\prime}, \omega +\Omega  }^{\rm{r}} 
     n_{\rm{L}}(\omega ) [1+n_{\rm{R}}(\omega +\Omega )] {\Big{\} }}.   \nonumber
%%%%%%%%%%%%%%%%%%%%%%%%%%%%%%%%%%
\end{eqnarray}
\end{subequations}
%%%%%%%%%%%%%%%%%%%%%%%%%%%%%%
%%%%%%%%%%%%%%%%%%%%%%%%%%%%%%
See Ref. [\onlinecite{KNmagnonNoiseJunction}] for more detailed analytical results in the vicinity of $\Omega = 0$.
%those reduce to Ref. [\onlinecite{KNmagnonNoiseJunction}] in the dc-limit $\Omega = 0$.

%%%%%%%%%%%%%%%%%%%%%%%%%%%%%%
%%%%%%%%%%%%%%%%%%%%%%%%%%%%%%
\section{Asymmetric quantum shot noise}
\label{sec:4}
%%%%%%%%%%%%%%%%%%%%%%%%%%%%%%
%%%%%%%%%%%%%%%%%%%%%%%%%%%%%%

\begin{figure}[h]
\begin{center}
\includegraphics[width=7.2cm,clip]{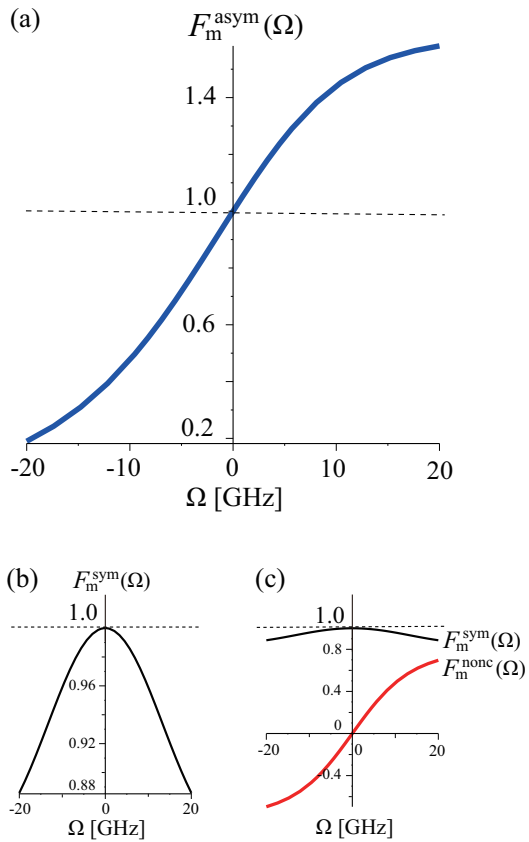}
\caption{
Plots of the frequency-dependent noise-to-current ratio in the magnonic shot noise regime at low temperatures as a function of $\Omega $ 
obtained by numerically solving Eq. (\ref{eqn:Sm2Omega}) for 
%%%%%%%%%%%%%%%%%%%%%%%%%%%%%%%%%%%%%%%%%%%%%%
$B_{\rm{L}} =  B_{\rm{R}} = 1 $T, 
$T_{\rm{L}} =  100 $mK, $T_{\rm{R}} =  190 $mK,
$ \eta = 9\times 10^{-2} $, and
$  0 \leq   \mid  \Omega  \mid   \leq  20  $GHz
with assuming a quadratic dispersion of magnons.
%%%%%%%%%%%%%%%%%%%%%%%%%%%%%%%%%%%%%%%%%%%%%%
%Each Fano factor does not depend on $J_{\rm{ex}}$, nor the spin stiffness constant.
%%%%%%%%%%%%%%%%%%%%%%%%%%%%%%%%%%%%%%%%%%%%%%
Panel (a): The noise-to-current ratio for the asymmetric noise 
$F^{\rm{asym}}_{\rm{m}} (\Omega )=F^{\rm{sym}}_{\rm{m}} (\Omega ) + F^{\rm{nonc}}_{\rm{m}} (\Omega )$. 
The noise-to-current ratio exceeds the constant `$1$' and the asymmetric shot noise becomes $ F^{\rm{asym}}_{\rm{m}} > 1  $ for $  \Omega  >0$.
%%%%%%%%%%%%%%%%%%%%%%%%%%%%%%%%%%%%%%%%%%%%%%
Panel (b): The noise-to-current ratio for the symmetric noise $F^{\rm{sym}}_{\rm{m}} (\Omega )$, being 
$  F^{\rm{sym}}_{\rm{m}} (\Omega ) =  F^{\rm{sym}}_{\rm{m}} (- \Omega )  $.
The symmetric shot noise exhibits $F^{\rm{sym}}_{\rm{m}} =1$ for $\Omega =0$ while $F^{\rm{sym}}_{\rm{m}} <1$ for $\Omega \not=0$.
The noise-to-current ratio does not exceed the constant `$1$'.
%%%%%%%%%%%%%%%%%%%%%%%%%%%%%%%%%%%%%%%%%%%%%%
Panel (c): The noise-to-current ratio for the noncommutative noise $ F^{\rm{nonc}}_{\rm{m}} (\Omega )  $. 
The noncommutative shot noise vanishes in the dc-limit $\Omega =0$ and thereby $F^{\rm{nonc}}_{\rm{m}} (\Omega =0)=0 $.
The noise-to-current ratio $ F^{\rm{nonc}}_{\rm{m}} (\Omega )  $ rapidly increases as a function of $\Omega $ monotonically, 
which is in contrast to the one for the symmetric noise $F^{\rm{sym}}_{\rm{m}} (\Omega )$.
%%%%%%%%%%%%%%%%%%%%%%%%%%%%%%%%%%%%%%%%%%%%%%
\label{fig:Noise3}
}
\end{center}
\end{figure}

To extract quantum-mechanical spin transport properties of magnons,
we now focus on the frequency-dependent noise spectrum for the spin current $ {\cal{I}}_{\rm{m}} $.
Introducing a magnonic analog of noise-to-current ratio for electron transport \cite{DLsuper-Poisson,DLsuper-Poisson2} by
%%%%%%%%%%%%%%%%%%%%%%%%%%%%%%%%%%%%%%%%
\begin{subequations}
\begin{eqnarray}
 F^{\rm{asym}}_{\rm{m}} (\Omega ) & \equiv  &  \frac{{\cal{S}}_{\rm{m}}^{\rm{asym}}(\Omega )}{g \mu _{\rm{B}} \mid {\cal{I}}_{\rm{m}} \mid }
=   \frac{2{\cal{S}}_{{\rm{m}}}(\Omega )}{g \mu _{\rm{B}} \mid {\cal{I}}_{\rm{m}} \mid},  \label{eqn:Fano1}   \\
%%%%%%%%%%%%%%%%%%%%%%%%%%%%%%%%%%%%%%%%%%%%%%%%
 F^{\rm{sym}}_{\rm{m}} (\Omega ) & \equiv  &  \frac{{\cal{S}}_{\rm{m}}^{\rm{sym}}(\Omega )}{g \mu _{\rm{B}} \mid {\cal{I}}_{\rm{m}} \mid }  
 =   \frac{{\cal{S}}_{{\rm{m}}}(\Omega ) + {\cal{S}}_{{\rm{m}}}(-\Omega )  }{g \mu _{\rm{B}} \mid {\cal{I}}_{\rm{m}} \mid}, \label{eqn:Fano2}    \\
 %%%%%%%%%%%%%%%%%%%%%%%%%%%%%%%%%%%%%%%%%%%%%%%%
 F^{\rm{nonc}}_{\rm{m}} (\Omega ) & \equiv  &  \frac{{\cal{S}}_{\rm{m}}^{\rm{nonc}}(\Omega )}{g \mu _{\rm{B}} \mid {\cal{I}}_{\rm{m}} \mid }
 =  \frac{{\cal{S}}_{{\rm{m}}}(\Omega ) - {\cal{S}}_{{\rm{m}}}(-\Omega )  }{g \mu _{\rm{B}} \mid {\cal{I}}_{\rm{m}} \mid},  
\label{eqn:Fano3}
\end{eqnarray}
\end{subequations}
we refer to $F^{\rm{asym}}_{\rm{m}} (\Omega )$, $F^{\rm{sym}}_{\rm{m}} (\Omega ) $, and $F^{\rm{nonc}}_{\rm{m}} (\Omega )  $ as 
the frequency-dependent noise-to-current ratio for the asymmetric, symmetric, and noncommutative noise of magnons, respectively.
%%%%%%%%%%%%%%%%%%%%%%%%%%%%%%%%
Focusing on a low frequency regime (i.e., a strong magnetic field)
\begin{eqnarray}
g \mu_{\rm{B}} B_{\rm{L(R)}}   \gg    \mid \hbar \Omega  \mid
\label{eqn:FrequencyScale}
\end{eqnarray}
with $B_{\rm{L}} = B_{\rm{R}}\equiv B$, we consider the magnonic shot noise region \footnote{Note that the magnonic shot noise regime does hold even for $   {\mid \Delta T\mid }/{T}  \geq 1   $ 
since $  \partial n_{\rm{R}}(T +  \Delta T)/\partial  \Delta T >0 $.}
\begin{eqnarray}
 {\mid  \Delta T \mid  }/{T}   \gg    {2 k_{\rm{B}}T}/{g \mu _{\rm{B}}B}
 \label{eqn:shotnoisecondition}
\end{eqnarray}
at low temperatures 
\begin{eqnarray}
{ k_{\rm{B}}T}/{g \mu _{\rm{B}}B}    \ll   1,
\label{eqn:shotnoisecondition2}
\end{eqnarray}
where equilibrium noise (i.e., thermal noise \cite{ThermalNoise,ThermalNoise2}) is strongly suppressed and 
nonequilibrium noise induced by the temperature difference $  \Delta T$ becomes dominant \cite{KNmagnonNoiseJunction}.
%%%%%%%%%%%%%%%%%%%%%%%%%%%%%%%%
The behavior of the frequency-dependent noise-to-current ratio in the magnonic shot noise regime is plotted in Fig. \ref{fig:Noise3}.
%%%%%%%%%%%%%%%%%%%%%%%%%%%%%%%%\cite{DLsuper-Poisson}

From Fig. \ref{fig:Noise3}
we see that the noncommutative noise associated with quantum fluctuations of magnon currents realizes 
the asymmetric `quantum' shot noise \cite{DLsuper-Poisson}
$ F^{\rm{asym}}_{\rm{m}} >1 $ [Fig. \ref{fig:Noise3} (a)];
the asymmetric noise consists of the symmetric noise and the noncommutative noise 
$  {\cal{S}}^{\rm{asym}}_{{\rm{m}}}(\Omega )  = {\cal{S}}^{\rm{sym}}_{{\rm{m}}}(\Omega )+ {\cal{S}}^{\rm{nonc}}_{{\rm{m}}}(\Omega )$,
and thereby 
\begin{eqnarray}
  F^{\rm{asym}}_{\rm{m}} (\Omega )  =   F^{\rm{sym}}_{\rm{m}} (\Omega ) + F^{\rm{nonc}}_{\rm{m}} (\Omega ). 
  \label{eqn:FanoSum}
\end{eqnarray}
%%%%%%%%%%%%%%%%%%%%%%%%%%%%%%%%%%%%%%%%%%%%%%%%%%%%%%%%%%%%
In the dc-limit $ \Omega =0$ the asymmetric noise becomes identical to the symmetric one 
 $ {\cal{S}}^{\rm{asym}}_{{\rm{m}}}(\Omega =0) = {\cal{S}}^{\rm{sym}}_{{\rm{m}}}(\Omega =0) $
since the noncommutative noise vanishes $  {\cal{S}}^{\rm{nonc}}_{{\rm{m}}}(\Omega =0) = 0  $.
%%%%%%%%%%%%%%%%%%%%%%%%%%%%%%%%%%%%%%%%%%%%%%%%%%%%%%%%%%%%
In that sense, the symmetric noise ${\cal{S}}^{\rm{sym}}_{{\rm{m}}}(\Omega =0)   $ in the dc-limit $ \Omega =0$ can be identified with `classical' noise.
There the noise-to-current ratio for the symmetric noise takes the maximum value [Fig. \ref{fig:Noise3} (b)]
\begin{eqnarray}
     F^{\rm{sym}}_{\rm{m}}(\Omega =0) =1,               
\end{eqnarray}
while it cannot exceed the constant `$1$' for any $\Omega $ 
\begin{eqnarray}
              F^{\rm{sym}}_{\rm{m}} (\Omega ) \leq 1
\end{eqnarray}
and the symmetric shot noise remains $F^{\rm{sym}}_{\rm{m}} (\Omega ) < 1$ for $\Omega \not=0$.
This means that $F^{\rm{sym}}_{\rm{m}} (\Omega ) = 1$ is the `classical' upper limit.
%%%%%%%%%%%%%%%%%%%%%%%%%%%%%%%%%%%%%%%%%%%%%%%%%%%%%%%%%%%%
However, the noncommutative noise, vanished in the dc-limit, rapidly increases as a function of $\Omega $ monotonically [Fig. \ref{fig:Noise3} (c)]
and consequently, the asymmetric shot noise becomes [Fig. \ref{fig:Noise3} (a)]
$ F^{\rm{asym}}_{\rm{m}} (\Omega ) =   F^{\rm{sym}}_{\rm{m}} (\Omega ) + F^{\rm{nonc}}_{\rm{m}} (\Omega ) >1 $ for $\Omega >0$.
Thus the noncommutative noise associated directly with quantum fluctuations of magnon currents breaks through the `classical' limit
and realizes the asymmetric `quantum' shot noise 
\begin{eqnarray}
           F^{\rm{asym}}_{\rm{m}} >1.
\end{eqnarray}
%%%%%%%%%%%%%%%%%%%%%%%%%%%%%%%%%%%%%%%%%%%%%%%%%%%%%%%%%%%%

%Note that the asymmetric shot noise above-mentioned is the fully quantum transport phenomenon which cannot be obtained by the LLG equation-based framework; the noncommutative noise vanishes in the classical regime where spin operators are identified with commutative magnetization vectors that obey the LLG equation. 

%%%%%%%%%%%%%%%%%%%%%%%%%%%%%
%%%%%%%%%%%%%%%%%%%%%%%%%%%%%
%%%%%%%%%%%%%%%%%%%%%%%%%
%%%%%%%%%%%%%%%%%%%%%%%%%

%%%%%%%%%%%%%%
%We optimistically expect that Nitrogen-vacancy center \cite{NVreview} or the muon spin rotation and relaxation \cite{muSR,muSR2} might be one of the most promising strategy toward observation and that, while being  still challenging, asymmetric quantum shot noise of magnons could become within experimental reach by developing those measurement schemes.
%Thus, we find that the quantum shot noise is super-Poissonian
%resulting in a large spin-Fano factor and super-Poisson spin-shot noise 

%%%%%%%%%%%%%%%%%%%%%%%%%%%%%%%%%
%%%%%%%%%%%%%%%%%%%%%%%%%%%%%%%%%
\section{Theoretical insights into experiment}
\label{sec:TheoreticalInsight}
%%%%%%%%%%%%%%%%%%%%%%%%%%%%%%%%%
%%%%%%%%%%%%%%%%%%%%%%%%%%%%%%%%%

Within the above theoretical studies using the junction model,
it will be instructive \cite{BalandinPrivate} to provide some insights into experiment.
%%%%%%%%%%%%%%%%%%%%%%%%%%%%%%%%%
Note that magnonic current-noise is now a measurable physical quantity and 
even the frequency-dependence is within experimental reach;
magnonic current-noise in an insulating ferromagnet has been measured at room temperature as a function of frequency 
in Ref. [\onlinecite{magnonNoiseMeasurement}], see Fig. 3 (a) where they have plotted the frequency-dependent magnonic current-noise
in the regime 
${\cal{O}}(10) {\rm{Hz}} - {\cal{O}}(10) {\rm{KHz}}$,
which is within our theoretical studies (Fig. \ref{fig:Noise3}).
%%%%%%%%%%%%%%%%%%%%%%%%%%%%%%%%%%%%%%%

%%%%%%%%%%%%%%%%%%%%%%%%%%%%%%%
\subsection{Experiment by Rumyantsev et al.}
\label{subsec:agreementExp}
%%%%%%%%%%%%%%%%%%%%%%%%%%%%%%%

%%%%%%%%%%%%%%%%%%%%%%%%%%%%%%%%%%%%
The frequency-dependence of the magnonic current-noise measured in the experiment
by Rumyantsev et al. \cite{magnonNoiseMeasurement}
[Fig. 3 (a) of Ref. [\onlinecite{magnonNoiseMeasurement}]]
shows good agreement \cite{BalandinPrivate}
with that of our theoretical studies on symmetric noise [Fig. \ref{fig:Noise3} (b)];
%%%%%%%%%%%%%%%%%%%%%%%%%%%%%%%%%%%%
from Figs. \ref{fig:Noise3} (a) and (b) for $\Omega  > 0$
we see that asymmetric noise increases as a function of frequency, 
while symmetric noise decreases. 
%%%%%%%%%%%%%%%%%%%%%%%%%%%%%%%%%%%%%%%%%%%%%%%%%%%%%
Since the magnonic current-noise of the experiment \cite{magnonNoiseMeasurement}
decreases as a function of frequency,
we identify the noise measured in Ref. [\onlinecite{magnonNoiseMeasurement}]
with symmetric noise.
%%%%%%%%%%%%%%%%%%%%%%%%%%%%%%%%%%%%%%%%%%%%%%%%%%%%%
This is our theoretical insight into the experiment by Rumyantsev et al \cite{magnonNoiseMeasurement}.

We remark that according to Ref. [\onlinecite{magnonNoiseMeasurement}],
their experimental scheme ensures that the background noise is negligibly small, i.e.,
the level of the background noise is at least an order of magnitude smaller than that measured magnonic current-noise.

%%%%%%%%%%%%%%%%%%%%%%%%%%%%%
%%%%%%%%%%%%%%%%%%%%%%%%%%%%%
\subsection{Measurement scheme for observation of asymmetric noise in magnon transport}
\label{subsec:ObservationAsymmetric}
%%%%%%%%%%%%%%%%%%%%%%%%%%%%%
%%%%%%%%%%%%%%%%%%%%%%%%%%%%%

On top of it, according to Fig. 3 (b) of Ref. [\onlinecite{magnonNoiseMeasurement}],
magnonic current-noise is measurable also as a function of time.
The Fourier transform of the measured time-dependent noise corresponds to the noise spectrum 
$ {\cal{S}}_{{\rm{m}}}(\Omega )  $ of Eq. (\ref{eqn:DefNoiseSpectrumKKK}) in our theory,
and asymmetric noise spectrum is given by $ {\cal{S}}_{{\rm{m}}}(\Omega )  $ itself,
i.e., $  {\cal{S}}^{\rm{asym}}_{{\rm{m}}}(\Omega )   =2 {\cal{S}}_{{\rm{m}}}(\Omega ) $.
%%%%%%%%%%%%%%%%%%%%%%%%%%%%%%%%%%%%%%%%%%%%%%%%%%%%%%
Those [Figs. 3 (a) and (b) of Ref. [\onlinecite{magnonNoiseMeasurement}]]
mean that there are no technical obstacles to the observation of asymmetric noise in magnon transport;
%%%%%%%%%%%%%%%%%%%%%%%%%%%%%%%%%%%%%%%%%%%%%%%%%%%%%%
making use of the measurement techniques \cite{magnonNoiseMeasurement},
the asymmetric noise is measurable and now within experimental reach with current device and measurement scheme.
%%%%%%%%%%%%%%%%%%%%%%%%%%%%%%%%%%%%%%%%%%%%%%%%%%%%%%

However, we stress that low temperature is required.
Low temperature is essential to the observation of asymmetric noise in magnon transport
instead of room temperature \cite{magnonNoiseMeasurement}
since noncommutative noise [Fig. \ref{fig:Noise3} (c)] associated directly with quantum fluctuations of magnon currents
and with the resulting asymmetry is essentially and truly quantum effect.
%%%%%%%%%%%%%%%%%%%%%%%%%%%%%%%%%%%%%%%
Moreover, the temperature scale is given by the magnon gap for magnons [Eq. (\ref{eqn:shotnoisecondition2})]
(e.g., the Zeeman energy $ g \mu _{\rm{B}}B/k_{\rm{B}} \sim  1 $K),
while by the Fermi temperature ($\sim  10^4  $K for normal metals) for electrons.
%%%%%%%%%%%%%%%%%%%%%%%%%%%%%%%%%%%%%%%
Therefore low temperature is required for the observation of asymmetric noise in magnon transport
instead of room temperature \cite{magnonNoiseMeasurement}.
%%%%%%%%%%%%%%%%%%%%%%%%%%%%%%%%%%%%%%%
This is the difference from electron systems;
the low temperature is the key ingredient for the observation of asymmetric noise in magnon transport.

%%%%%%%%%%%%%%%%%%%%%%%%%%%%%%%%%%%%%%%
For an estimate, we assume the following experiment parameter values 
$ B = 1$T,
$ \Delta T   =1$mK,
$T=10$mK.
%%%%%%%%%%%%%%%%%%%%%%%%%%%%%%%%%%%%%%%
Since the frequency scale is given by the magnon gap, e.g., the Zeeman energy [Eq. (\ref{eqn:FrequencyScale})],
the low frequency regime appropriate to the observation becomes 
$  0 \leq   \mid  \Omega  \mid   \leq  20  $GHz.
%%%%%%%%%%%%%%%%%%%%%%%%%%%%%%%%%%%%%%%
Note that at such low temperatures [Eq. (\ref{eqn:shotnoisecondition2})], i.e., $ { k_{\rm{B}}T}/{g \mu _{\rm{B}}B}    \ll   1 $,
shot noise properties do not depend on details of the magnon dispersion \cite{KNmagnonNoiseJunction}, and 
interaction effects (e.g., magnon-magnon and magnon-phonon interactions)
are assumed to be negligibly small \cite{adachiphonon,Tmagnonphonon}.
%%%%%%%%%%%%%%%%%%%%%%%%%%%%%%%%%%%%%%%
Moreover, the measurement scheme \cite{magnonNoiseMeasurement}
ensures that the background noise is suppressed and negligibly small, i.e., the level of the background noise is at least an order of magnitude smaller than that measured magnonic current-noise.
%%%%%%%%%%%%%%%%%%%%%%%%%%%%%%%%%%%%%%%
Given these estimates,
we conclude that at low temperature instead of room temperature 
the observation of asymmetric noise in magnon transport, 
while being challenging, 
seems within experimental reach with current device and measurement techniques \cite{magnonNoiseMeasurement}.
%%%%%%%%%%%%%%%%%%%%%%%%%%%%%%%%%%%%%%%

Making use of the above measurement scheme while exploiting low temperatures, 
asymmetric noise in magnon transport will be observed.
Lastly, we predict that, in contrast to the one of quantum dots \cite{DLsuper-Poisson},
there will be no dips near the zero-frequency regime in the observed asymmetric noise of magnon currents.
%%%%%%%%%%%%%%%%%%%
The dip theoretically predicted in Ref. [\onlinecite{DLsuper-Poisson}]
is due to the charging effects of the dot and it is intrinsic to the system, quantum dots,
consisting of electric charge degrees of freedom.
%%%%%%%%%%%%%%%%%%%
%%%%%%%%%%%%%%%%%%%
We thus predict the difference between asymmetric noise of magnons in insulating ferromagnets [Fig. \ref{fig:Noise3} (a)]
and that of electrons in quantum dots, i.e., 
the absence/presence of dips near the zero-frequency regime;
the frequency-dependence of asymmetric noise varies from system to system, e.g., 
depending on whether it consists of spin or electric charge degrees of freedom.
%%%%%%%%%%%%%%%%%%%

%%%%%%%%%%%%%%%%%
%%%%%%%%%%%%%%%%%
\section{Summary}
\label{sec:conclusion}
%%%%%%%%%%%%%%%%%
%%%%%%%%%%%%%%%%%

We have studied asymmetric quantum shot noise in magnon transport 
and determined the frequency dependence of noise-to-current ratio for asymmetric, symmetric, and noncommutative noise at low temperatures.
%%%%%%%%%%%%%%%%%%%%%%%%%%%%%%%%%%%%%%
We found that the noncommutative noise associated directly with quantum fluctuations of magnon currents breaks through the `classical' limit of the ratio determined by the symmetric noise and realizes the asymmetric `quantum' shot noise.
%%%%%%%%%%%%%%%%%%%%%%%%%%%%%%%%%%%%%%
As seen in the Heisenberg's uncertainty relation, the noncommutativity of operators lies at the heart of quantum mechanics;
in that sense, the asymmetric quantum shot noise assisted by the noncommutative noise is identified with fully `quantum' transport phenomenon. 
%in the classical regime where the noncommutativity of spin operators ceases to work and spin operators can be treated as commutative magnetization vectors, the noncommutative noise vanishes.
%which cannot be obtained by the LLG equation-based description.
%%%%%%%%%%%%%%%%%%%%%%%%%%%%%%%%%%%%%%

Using the magnon description instead of the LLG description, one can take into account quantum fluctuations of `magnons'.
%However, the purpose of this paper is not quantum fluctuations of `magnons', but quantum fluctuations of `magnon current'.
However, we stress that the magnon description does not ensure quantum fluctuations of `magnon currents'.
Note that while quantum fluctuations of `magnons' are included into the symmetric noise,
quantum fluctuations of `magnon currents' are not included into the symmetric noise.
%%%%%%%%%%%%%%%%%%%%%%%%%%%%%%%%%%
On the other hand,
quantum fluctuations of `magnons' and those of `magnon currents' are included into the asymmetric noise.
%%%%%%%%%%%%%%%%%%%%%%%%%%%%%%%%%%
We have thus seen that even if one employs the magnon description,
it does not mean quantum fluctuations of `magnon currents' are included into the theory.
%%%%%%%%%%%%%%%%%%%%%%%%%%%%%%%%%%
In this paper exploiting magnonic current-noise defined as the current-current correlation function,
we have thus demonstrated that
even within the magnon description there still exists a class of `classical vs quantum', i.e.,
symmetric shot noise of magnons and asymmetric shot noise of magnons, respectively;
asymmetric shot noise is identified with a `quantum' phenomenon,
while symmetric shot noise is identified with a `classical' one (compared with asymmetric noise)
in the sense that quantum fluctuations of `magnon currents' are not included into the symmetric noise.
%%%%%%%%%%%%%%%%%%%%%%%%%%%%%%%%%%%%%%%%%%%%%%%%%%
The noncommutative noise associated directly with quantum fluctuations of `magnon currents' thus plays a significant role 
in distinguishing the class of `classical vs quantum' in the magnon description, i.e.,  
symmetric shot noise of magnons and asymmetric shot noise of magnons, respectively.
%%%%%%%%%%%%%%%%%%%%%%%%%%%%%%%%%%%%%%%%%%%%%%%%%%

Those results of this paper has been obtained by focusing on the `current-current' correlation function of magnons (i.e., noise) 
instead of `current' itself; the central issue of spintronics research \cite{mod2} so far is spin current, while recently magnonic current-noise has been measured as a function of frequency  in an insulating ferromagnet \cite{magnonNoiseMeasurement}.
%%%%%%%%%%%%%%%%%%%%%%%%%%%%%%%%
Those highlight that 
our work exploiting the current-current correlation function of magnons 
goes beyond the current-based spintronics
and provides a new direction to magnon-based spintronics \cite{MagnonSpintronics}, quantum magnonics;
%%%%%%%%%%%%%%%%%%%%%%%%%%%%%%%%%%%%%%
making use of the recently reported measurement scheme \cite{magnonNoiseMeasurement} for the frequency-dependent magnonic current-noise,
our theoretical predictions are within experimental reach with current device and measurement technologies while exploiting low temperatures.
%%%%%%%%%%%%%%%%%%%%%%%%%%%%%%%%%%%%%%
Our work thus provides a platform toward experimental access to quantum fluctuations of magnon currents.
We believe this work serves as a bridge between two research areas, spintronics and mesoscopic physics.
%%%%%%%%%%%%%%%%%%%%%%%%%%%%%%%%%%%%%%

%%%%%%%%%%%%%%%%%
%%%%%%%%%%%%%%%%%
\section{Discussion}
\label{sec:discussion}
%%%%%%%%%%%%%%%%%
%%%%%%%%%%%%%%%%%

Within the present our theoretical studies
we have provided several insights into the experiment by Rumyantsev et al. \cite{magnonNoiseMeasurement};
while their measured magnonic current-noise is identified with symmetric noise,
the asymmetric noise is now within experimental reach with current device and their measurement technologies
by exploiting low temperatures.
%%%%%%%%%%%%%%%%%%%%%%%%%%%%%%%%
Still, a general treatment \cite{Altshuler} of magnonic current-noise in insulating magnets, 
going beyond the present our theoretical framework,
deserves further study,
and some of them will be addressed elsewhere in the near future.
%%%%%%%%%%%%%%%%%%%%%%%%%%%%%%%%

See Refs. [\onlinecite{Kamra2,Kamra3}] for squeezing of magnons and quantum contributions to symmetric noise of spin currents in ferromagnet/nonmagnetic conductor hybrids subjected to a coherent microwave drive;
where following a mathematical analogy with current noise for electrons,
a physical understanding of the frequency dependence for a microwave-induced symmetric spin current noise 
and the quantum contribution are discussed in terms of a photon-like quasiparticle picture.
%%%%%%%%%%%%%%%%%%%%%%%%%%%%%%%%%%%%%%%%%%%%%%%%%%%%%%%%%%%%%
See Refs. [\onlinecite{Kamra2,Kamra3}] for details.
%%%%%%%%%%%%%%%%%%%%%%%%%%%%%%%%%%%%%%%%%%%%%%%%%%%%%%%%%%%%%

%%%%%%%%%%%%%%%%%%%%%%%%%%%%%%%%%%%%%%%%%%%%%%%%%%%%%%%%%%%%%
%%%%%%%%%%%%%%%
\begin{acknowledgments}
%%%%%%%%%%%%%%%
This work (KN) is supported by Leading Initiative for Excellent Young Researchers, MEXT, Japan.
We would like to thank D. Loss for stimulating discussion about asymmetric quantum shot noise,
H. Chudo for fruitful  discussion about experiment and data analysis,
and A. A. Balandin for helpful discussion about their measurement \cite{magnonNoiseMeasurement} with encouragement.
%%%%%%%%%%%%%%%%%%%%%%%%%%%%%%%%%%%%%%%%%%%%%%%%%%%%%%%%%%%%%
We also thank anonymous referees for the valuable comments that helped us improve the manuscript.
%%%%%%%%%%%%%%
\end{acknowledgments}
%%%%%%%%%%%%%%
%%%%%%%%%%%%%%%%%%%%%%%%%%%%%%%%%%%%%%%%%%%%%%%%%%%%%%%%%%%%%

\bibliography{PumpingRef}

\end{document}